\documentstyle[11pt,newpasp,twoside,epsf]{article}
\def\edcomment#1{\iffalse\marginpar{\raggedright\sl#1\/}\else\relax\fi}
\reversemarginpar

\begin{document}

\title{Giant Planets and Variable Stars in 47 Tucanae - a progress report of a comprehensive ground-based search.}

\author{David T F Weldrake, Penny D Sackett}

\affil{Research School of Astronomy and Astrophysics, Mount Stromlo Observatory, Cotter Road, Weston Creek, ACT 2611, Australia}

\author{Terry J Bridges}

\affil{Anglo-Australian Observatory, P.O. Box 296, Epping. NSW, 1710, Australia}

\begin{abstract}
Here we present a short progress report of a comprehensive search for variability in the globular cluster 47 Tucanae. Using the MSSSO 40'' telescope and a combined V+R filter, we are searching for variability across a 52x52' field centered on the cluster. The main aim is to search for transiting 'Hot Jupiter' planets in this cluster, the results of which are still being produced, but a natural side product is a deep catalogue of variable stars within the cluster field. The experiment samples the whole of the cluster (except the inner 5'), thus probing the uncrowded outer regions where the stellar densities are lower, increasing the prospects for the survivability of planetary systems. Half of the currently identified variable stars are new discoveries. We have data for 36,000 stars with masses similar to that of the Sun for the main transit search. 
\end{abstract}

\section{Introduction}
We are searching for planetary transits of main sequence (MS) stars in the globular cluster 47 Tucanae. Motivated by the detection of the transit of HD209458 (Charbonneau et al 2000), and also by the apparent lack of transit detections in the inner, crowded regions of 47Tuc (Gilliland et al 2000), we believe observing the less crowded outer regions of the cluster will help us properly determine the distribution and frequency of short-period giant planets (``Hot Jupiters'') in globular clusters. 47Tuc is an ideal target due to its proximity, observability throughout the night from Siding Spring, distance from the moon, and total mass (number of stars).

Transit observations allow us to determine a number of planetary characteristics, including the orbital period, an estimate of the orbital inclination and a direct measurement of the planet size. For a period of a few hours, a planet of Jupiter's size will cause a 1-2$\%$ dimming in a solar-type MS star. It would be very difficult to see a transit of a more evolved star, since the percentage dimming would be significantly smaller. Recent doppler periodicity observations have shown that 51-Peg-type 'Hot Jupiter' planets are more prevalent than previously thought, although this method is naturally biased towards detecting such planets. Nevertheless it is believed that they are present around 2-3$\%$ of solar mass stars (Cumming et al 1999). With the sheer number of such stars to observe in 47Tuc, our prospects for a successful detection are good, if the frequency of Hot Jupiters in the outer parts of 47Tuc is the same as that in the Solar Neighbourhood.

\section{Observations and Data Reduction}
Using the 40$''$ telescope at Siding Spring Observatory, we observed 47 Tuc for 33 nights in Aug/Sept 2001. We obtained a total of 1220 images of the cluster, each of 300s exposure, and used them to produce time-series lightcurves for all visible unsaturated stars of 15$\leq$V$\leq$21. This covers a large range of stellar type, encompassing the bottom of the Red Giant Branch (RGB), the Subgiant Branch, the cluster Turn-off and the top of the cluster MS.

We used the Wide Field Imager (WFI) which consists of a 4x2 array of 2048x4096 pixel back-illuminated CCDs, arranged to give a total format of 8Kx8K pixels. The CCDs give a scale of 0.38"/pix, with a field-of-view of 52$'$ on a side. We employed a combined Cousins V+R filter to achieve maximum signal-to-noise (SN). For our main aim of detecting planetary transits, we need a SN ratio of 200 at V=18.

Initial reductions were undertaken with the MSCRED package within IRAF. This included trim and overscan correction, bias correction, flat-field and dark current subtraction. The main photometric analysis was carried out using the Difference Image Analysis (DIA) method described by Alard $\&$ Lupton (1999), and modified by Wozniak (2000). This method has been successfully used by OGLEIII for variability searches in globular clusters (eg; Kaluzny et al 1998). 
\begin{figure}[!htbp]
\centerline{
\plottwo{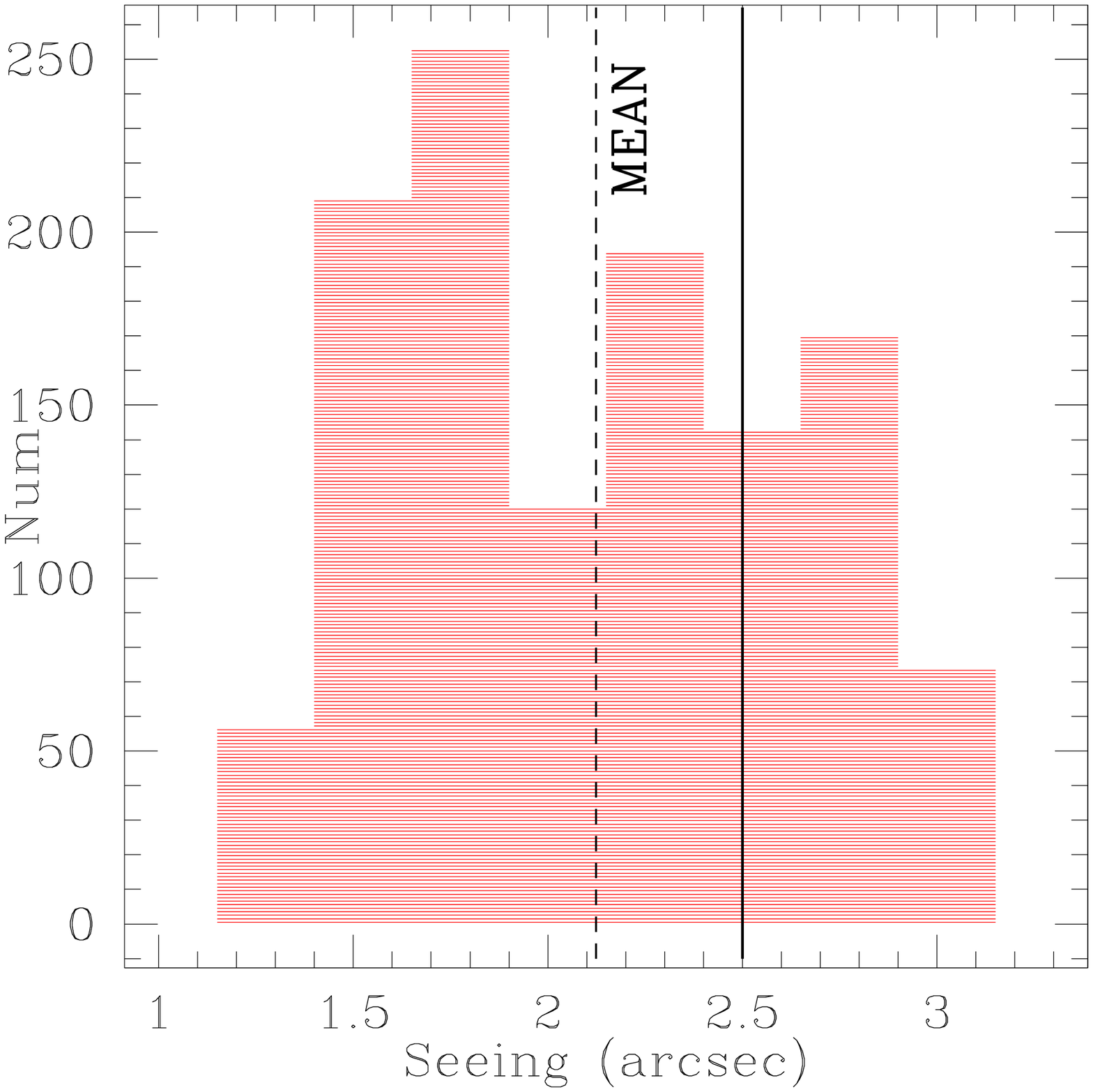}{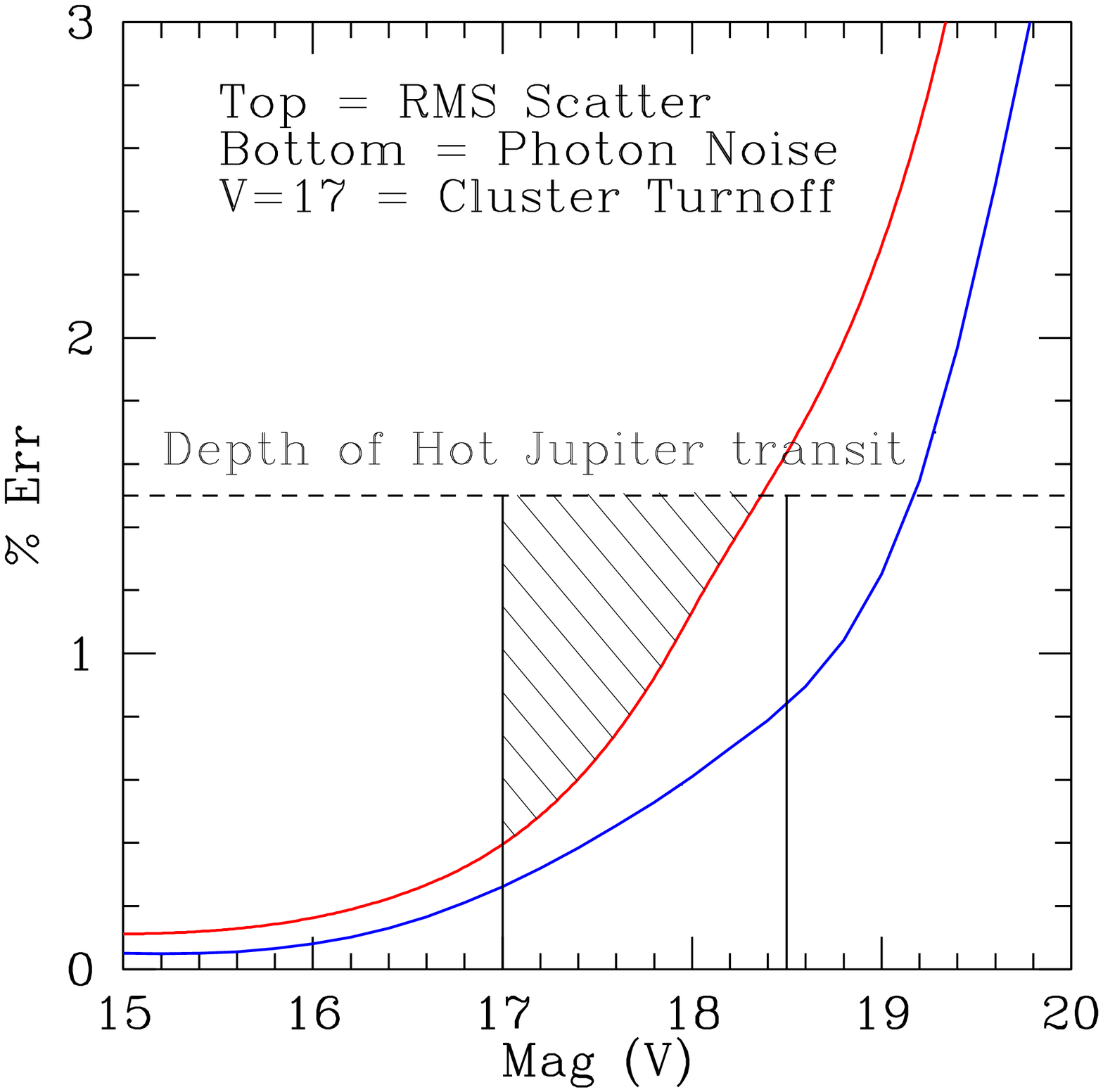}
}
\caption{Left: Histogram of seeing versus number of images. Seeing $\leq$2.5$''$ gives a sufficient SN for transit detection. Right: Photometric quality versus apparent magnitude, dashed line indicates depth of HD209458 transit. Box outlines our transit search range. RMS scatter is good enough to V=18.38 to detect a 1.5$\%$ dip (shaded region).}
\end{figure}

\section{Comparison to the HST 47Tuc Planet Search}
A recent 8-day experiment with the HST, in which 30,000 47Tuc stars (most of which are 3-4 mag below MS turn-off) were studied, found no transits. Around 15-20 planets were expected (Gilliland et al 2000), assuming the number frequency of Hot Jupiters is the same in 47Tuc as in the Solar Neighbourhood. The field observed was very close (0.6$'$) to the globular centre, where the dynamical environment is potentially destructive for proto-planetary disks. Also the target stars were much less massive than the Sun, allowing for easier transit detection. The overall conclusion of this study was that the number of Hot Jupiters in Globular Clusters is at least an order of magnitude less than in the Solar Neighbourhood. 

In contrast, our experiment probes 36,000 Main Sequence stars of 17.0$\leq$V $\leq$18.5, which are more like the solar-type stars in mass, around which Hot Jupiters are already known. The long time span and large field of view (52x52$'$) allows us to probe a wider range of periods over the whole face of the cluster. This taken together, allows a more even comparison of environments with the Solar Neighbourhood.

\section{Expected Numbers}
Our calculations below assume that the frequency of 47Tuc 'Hot Jupiters' is the same as that observed in the Solar Neighbourhood. As the geometric probability of a transit is inversely proportional to the planetary semi-major axis, short-period planets are much more likely to be detected. Each planet to which we are sensitive has a $\sim$7$\%$ probability of transiting, and so a large sample of stars is needed for a successful result. We estimate the number of planets expected taking into account:

\begin{itemize}
\item Number of appropriate stars: 36000 stars V=17.0 to V=18.5 within 52$'$ area, excluding inner 5$'$ radius. 
\item Present detection rate of Hot Jupiters by RV techniques in the Solar Neighbourhood (0.7$\%$ (Butler et al 2002)).
\item The probability that such a planet will transit (about 7$\%$).
\item Transit duration (2.5 hrs depending on inclination, period and stellar type).
\item Effects of stellar metallicity on transit depth and duration.
\item Most typical periods (3-5d) of Hot Jupiters, averaged over orbital phase.
\item Observing Duty cycle (1/2) $\&$ weather conditions ($~$80$\%$ usable).
\end{itemize}

With these assumptions, the number of planets we expect to cause detectable transits over the face of 47Tuc with 33 nights of observing is 17.6$\pm$4.2. A null result would hence have a high significance. 

\section{Light Curve Analysis 1: Periodogram}
As of 5 Aug 2003, a total of 53853 lightcurves for one half the cluster face have been obtained with the DIA method, of which 17191 are within the transit search magnitude range. A Lomb-Scargle Periodogram algorithm was modified to search for periodicities in this data. To date, this method has identified 34 variable stars in the 47Tuc field; 17 are new discoveries, and shall be published when the catalogue is completed (Weldrake et al, 2003, in progress).

\begin{figure}[!htbp]
\centerline{
\plottwo{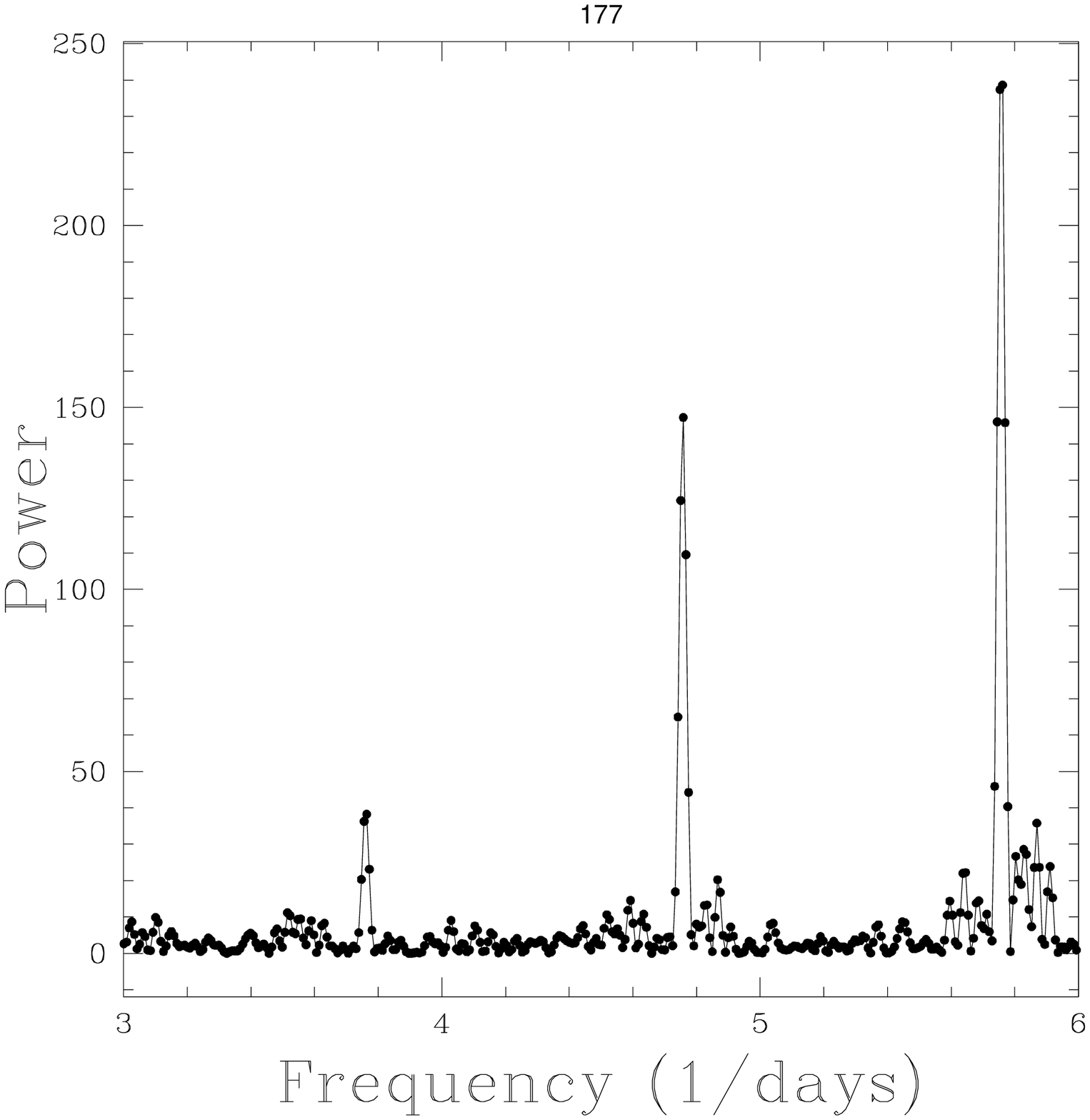}{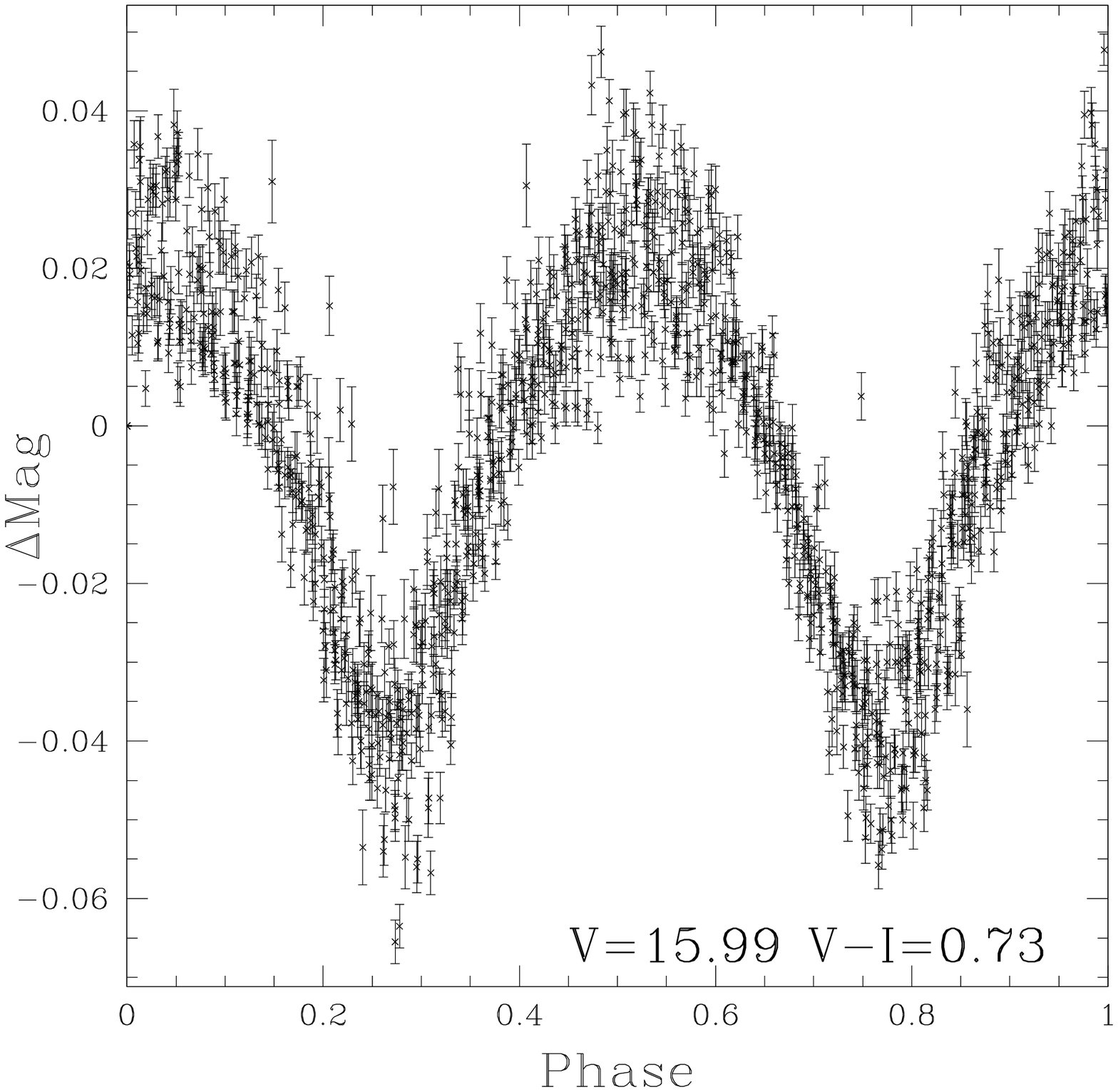}
}
\caption{Left: Example power spectrum; periodicities are well defined. Right: When the same lightcurve phase-wrapped to 0.347d, the lightcurve of a newly-discovered eclipsing binary is revealed. Note the very small amplitude. This is likely a foreground system.}
\end{figure}

\section{Light Curve Analysis 2: Matched Filter Algorithm}
To specifically search for planetary transit signatures, a matched filter algorithm (MFA) has been developed and will be applied to the transit search range lightcurves. This method compares the lightcurves to a model transit within a range of previously defined parameter space for transit depth, duration, period and MJD of the transit start. For a transit-present case, the output $\chi$$^2$ is significantly less than that of a no-transit-present case (figure 3).

\begin{figure}[!htbp]
\centerline{
\plottwo{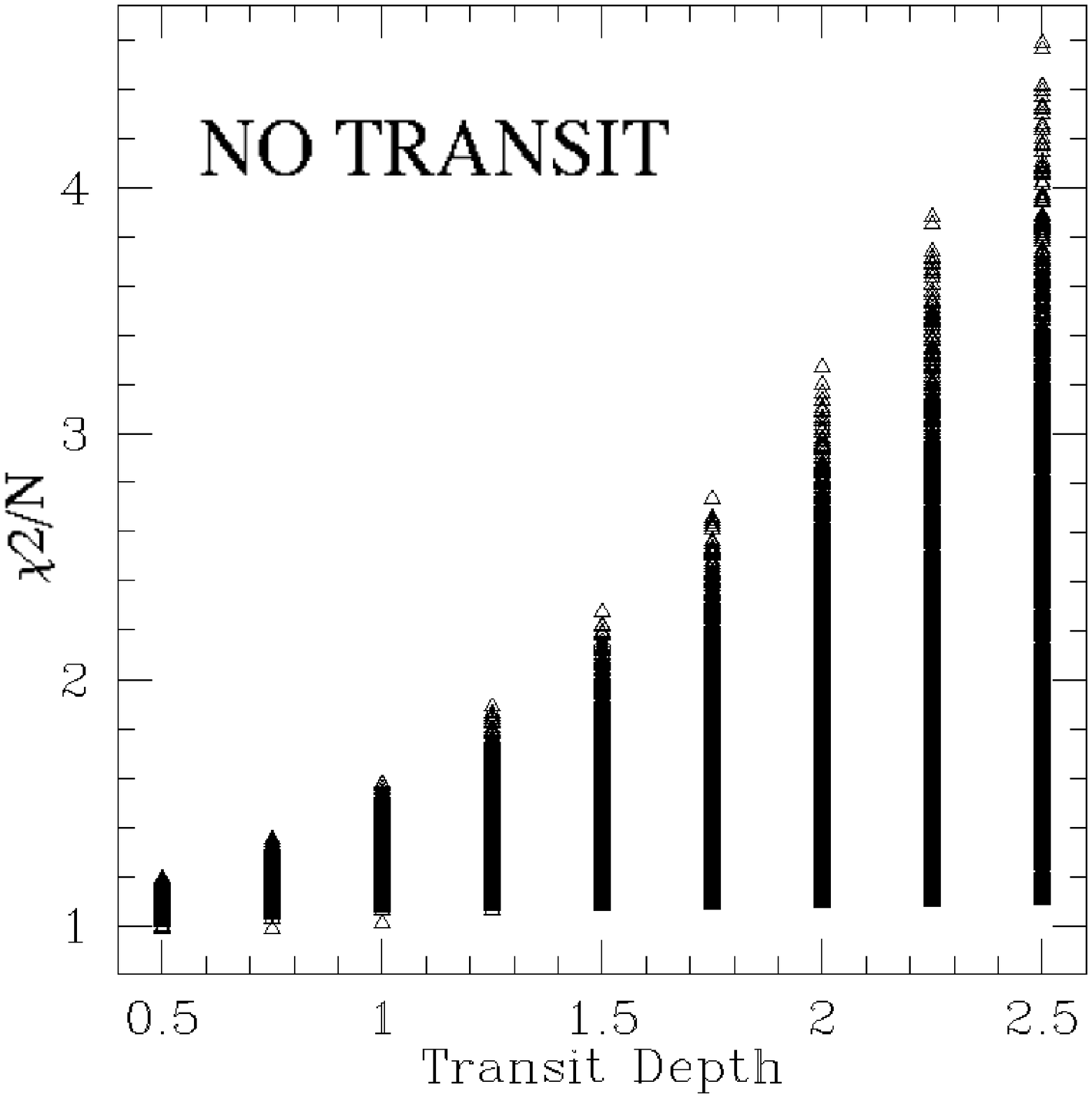}{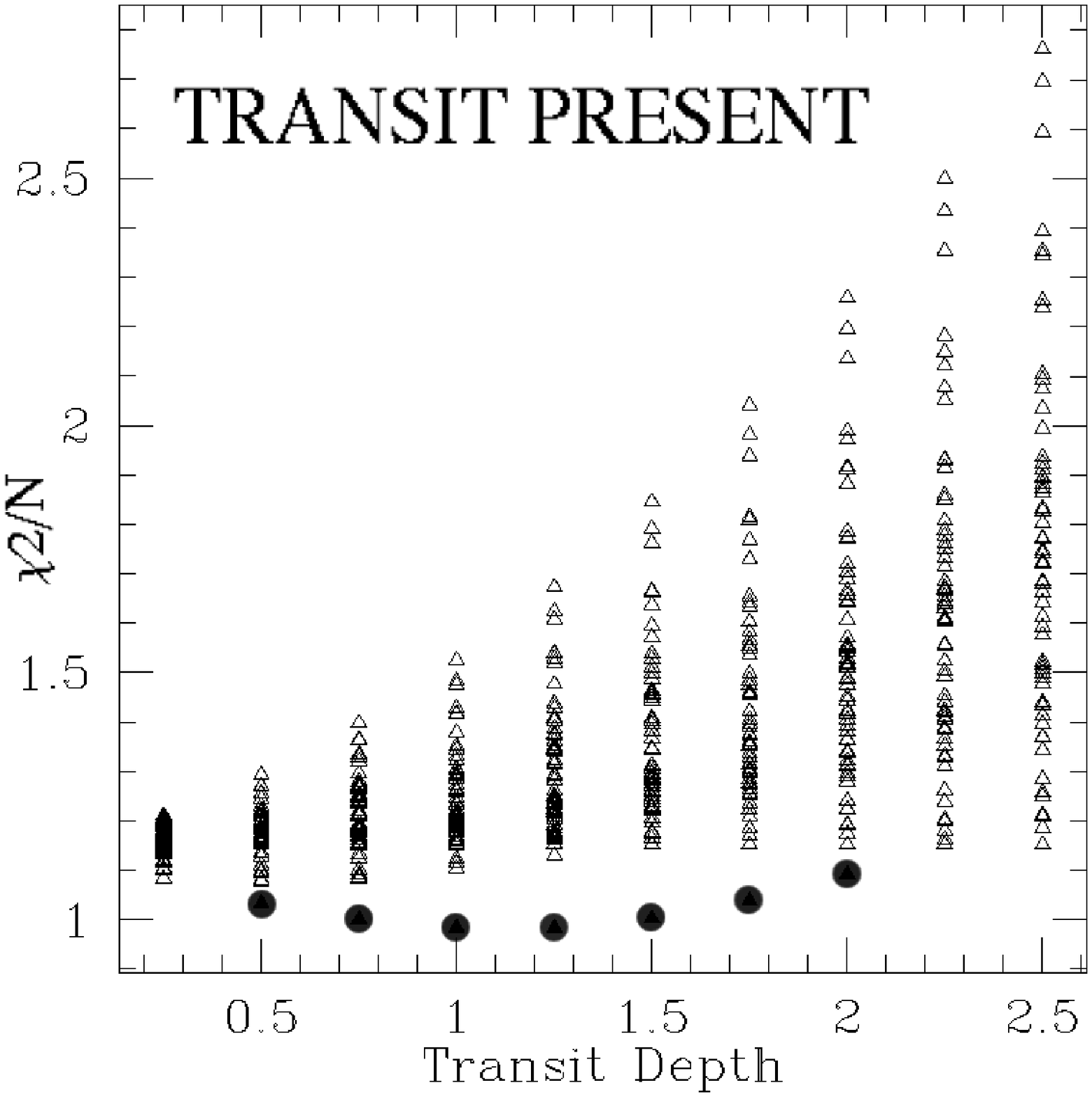}
}
\caption{Left: Search for transit in constant lightcurve, showing output $\chi$$^2$/N values for a large range of assumed model transit parameters. Right: Search applied to a V=18 lightcurve with a 1.5$\%$ transit introduced, lower $\chi$$^2$ values show a detection (filled circles). }
\end{figure}

\section{Results so far}
The variable star catalogue is well underway, with the identification of 34 variables, of which 17 are new discoveries. These variables cover a large range of types, including background SMC RR Lyraes, likely cluster Blue Stragglers, MS eclipsing binaries, 47Tuc pulsating Red Giants, SMC pulsating Red Giants, currently one Galactic Halo RR Lyrae and one likely SMC Blue Straggler.

\section{Summary}
We have presented a current progress report of a project to search for variability in the globular cluster 47Tuc, in particular a search for transiting 'Hot Jupiter' planets. In contrast to the Gilliland et al HST search, our work allows a more direct comparison of the number density of Hot Jupiters in 47Tuc to the Solar Neighbourhood. 

The periodogram analysis is completed for 1/4 of the field of 47Tuc, with the identification of 34 variable stars, 17 of which are new discoveries. The Matched Filter Algorithm is currently being tested and will be used to provide a more sensitive search for planetary transit signatures. The variable star catalogue is well underway, and shall be published in its entirety upon completion.

\end{document}